\documentclass[twocolumn,aps,showpacs,preprintnumbers,amsmath,amssymb,ltxgrid,rotate]{revtex4}
\usepackage{graphicx,color,epsfig}
\begin{document}
\bibliographystyle{apsrev}

\title[Charge effect on polymerization]{{\em Ab initio} study of charge doping effect on 1D polymerization of C$_{60}$}

\author{R. Poloni$^{1}$, A. San Miguel$^{2}$ and M. V. Fernandez-Serra$^{3}$}
\affiliation{
(1) Department of Chemical and Biomolecular Engineering, University of California, Berkeley 94720-1462, USA}
\affiliation{
(2)  Universit\'{e} Lyon 1,
Laboratoire PMCN; CNRS, UMR 5586; 69622 Villeurbanne Cedex, France}
\affiliation{
(3) State University of New York, Stony Brook, New York 11790, USA}

\preprint{version {\today}}
\date{\today}
\begin{abstract}
We study the interplay between charge doping and intermolecular distance in the polymerization of C$_{60}$ fullerene chains by means of density functional theory (DFT)-based first principle calculations.
	The potential energy surface analysis shows that both the equilibrium intermolecular distance of the unpolymerized system and the polymerization energy barrier are inversely proportional to the electron doping of the system. We analyze the origin of this charge-induced polymerization effect by studying the behavior of the system's wavefunctions around the Fermi level and the structural modifications of the molecules as a function of two variables: the distance between the centers of the molecules and the number of electrons added to the system.
\end{abstract}
\maketitle

\section{Introduction}
\label{intro}

The formation of covalent intermolecular bonds in solid C$_{60}$ has been observed to occur under a great variety of chemical and physical conditions.
Experimentally, the addition reaction between neighboring fullerenes was first observed upon visible or ultraviolet light treatment \cite{Rao1993} to crystals of pristine C$_{60}$. Later, the application of high pressure and high temperature conditions  \cite{Iwa1994,Nun1995,Sun1999} as well as electric field stimulated techniques \cite{Zhao1994}, have been reported as efficient methods for the formation of chains of polymerized fullerene molecules in solid C$_{60}$. 
The intercalation of alkali metal atoms in fullerite crystals has been observed to favor or prevent the formation of  C$_{60}$ intermolecular covalent bonds, depending on the nature of the intercalated atoms and on the stoichiometry of the crystal. Low content or low size of alkali atoms actually favor the formation of  C$_{60}$ polymers.
In particular, the MC$_{60}$ compounds (M=K, Rb and Cs)
undergo a structural transition from a  
high-temperature monomer to a polymer phase 
for T$<$350 K \cite{Pek1994,Ste1994,LauMor1998,Ver2002}. 
The C$_{60}$-based compounds intercalated
 with light alkali atoms (Li and Na), like Na$_2$RbC$_{60}$, Na$_4$C$_{60}$ and Li$_4$C$_{60}$
 have been observed to
spontaneously form 1D and 2D polymers, respectively,
at ambient conditions \cite{Dre1996,Mar2004}.
The polymeriazation between two neighboring molecules usually takes place via a [2+2]-cycloaddition. In fact, the forbiddness of the suprafacial [2+2]-cycloaddition, as stated by the Woodward-Hoffmann rule, is removed by the symmetry breaking of the orbital spectrum resulting from electron excitation, electron doping or a structural displacement occurring in the transition state  \cite{Emi2002}.
%
On the other hand, fullerides with high stoichiometries (e.g. M$_6$C$_{60}$, M=alkali metal) of light or heavy alkali atoms have not been observed to form C$_{60}$ polymers at ambient conditions of temperature and pressure. 
We recently showed that for high content of Rb or Cs metal atoms  a significant smaller distance between molecules, compared to pristine C$_{60}$, is required to observe a transition to a polymerized phase \cite{PolB2008,PolC2008,PolTou2010}.
 All this suggests that the interplay between electron doping and lattice parameter is essential to understand the polymerization.
Theoretical investigation about the stability of negatively charged isolated C$_{60}$ molecules \cite{Ped1992} and of anionic fullerenes chains connected through face-to-face hexagons have been already reported in earlier works \cite{SanLon2003}.
 Also, the electronic properties, geometric structure and optical properties of negatively charged dimers have been studied by employing semiempirical methods \cite{Kurti1996,Fage1996}.
More recently, the dimerization of C$_{60}$ has been studied in terms of the intermolecular interaction potential  
between isolated molecules by taking into account inter-ball donor-acceptor contributions \cite{She2007}.
In the present work, we perform a detailed study of the interplay between electron doping and inter-C$_{60}$ distance on the formation of inter-ball covalent bonds in attempt to understand the experimental results and guide future works. This knowledge is of importance to understand the interplay of charge 
transfer and steric effects on the polymerization of C$_{60}$ which can play 
a fundamental role in fullerene engineering of new materials for 
instance through high pressure and high temperature \cite{SanMiguel2006}.
This work reports a first principles calculations study based on the density functional theroy (DFT) on a 1D C$_{60}$ chain with explict addition electrons.

\section{Methodolody}
\label{methodology}
Calculations are performed with the \textsc{SIESTA} \cite{SolArt2002}
method, where the eigenstates of the Kohn-Sham hamiltonian
are expressed as linear combinations of numerical atomic
orbitals.
Many body effects are described within the local density approximation (LDA) to the exchange and correlation
potential \cite{PerZun1981}.
 We have used a
variationally optimized \cite{Jun2001,Ang2002} double-$\zeta$ polarized basis sets.
 Real space integrals were performed on a mesh with a 250 Ry cutoff.
Core electrons are replaced by  nonlocal, norm-conserving fully separable Trouiller-Martins pseudopotentials.
In the calculations  {\em 2s} and {\em 2p} electrons of C atoms were explicitly included in the valence. 
For each calculation, we have minimized the total energy of the system as a function of the atomic coordinates
imposing a tolerance in the largest component of the forces of 0.03 eV/\AA.
 A geometry optimization is performed 
for each intermolecular distance and for each charged state.

\begin{figure}[h]
\includegraphics[scale=0.2]{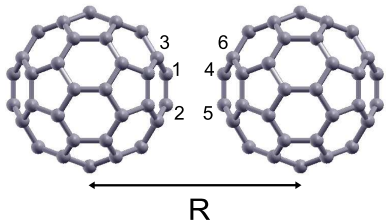}
\begin{centering}
\caption
{Arrangement of the molecules in the unit cell. $R$ is the center-to-center distance between neighboring fullerenes. Atoms numbered as 1, 2, 3 and 4 participate in the [2+2]-cycloaddition.}
\label{fig1}
\end{centering}
\end{figure}

Our study has been performed as a function of two variables:
(i) the {\em distance} $R$ (see Figure~\ref{fig1}) between the centers of the molecules  along the chain (uniformly varied from 8.50 \AA\ to 10.85 \AA);
(ii) the added {\em charge} {\em q}, consisting on the addition of 1, 2, 3 and 4 electrons to the system.
A neutralizing jellium background of opposite charge ensures system
neutrality as a whole in such a way that $\rho_{tot}(r)=\rho(r)+Q$, where $\rho_{tot}(r)$ is the electronic density of the system with the extra charge and $Q$ is the constant density background neutralizing the system \cite{LesGil1985,DeVGil1992}.

We employed a cubic supercell (of lattice parameter equal to $L$=$2R$), containing two molecules along one of the Cartesian axis in order to allow their relative reorientation during the study.
In this way, since the two nearest molecules at  $R$
 distance are placed
along the $<$001$>$ direction, the distance between two neighboring 
molecules along the  $<$010$>$ and the $<$100$>$ directions
 is large enough to allow us to neglect their interaction.
 We therefore refer to the studied systems as fullerene chains.

The starting relative-orientation of the molecules
 corresponds to that most frequently observed in 
the one-dimensional polymerization of 
 solid fcc C$_{60}$~\cite{Nun1995,MorLau1997},
 where a double bond of one molecule (atoms 1-2 in Figure~\ref{fig1}) opposes a double bond of the
 neighboring molecule (atoms 4-5 in Figure~\ref{fig1}).
 In this way, the polymerization takes place via a [2+2]-cycloaddition mechanism, meaning that two double bonds of two adjacent molecules break up in order to form two covalent bonds between the molecules. In the so formed four membered ring, each carbon atom is covalently bonded to other four C atoms through a mixed $sp^2$-$sp^3$ hybridized bond.

\section{Results and Discussion}
\subsection{Potential Energy surface}
\label{energy}
\begin{figure}[h]
\begin{centering}
\includegraphics[scale=0.7]{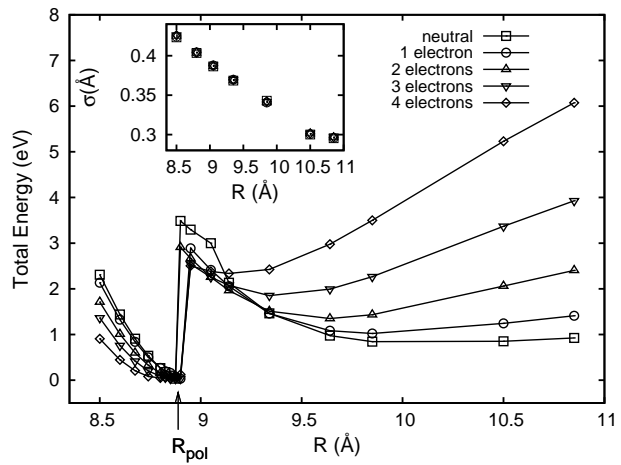}
\caption{Total energy profile as a function of the intermolecular  distance for different anionic states.
 Curves are shifted in such a way that the minimum energy value for each curve is set to zero. The energy values for points at $R>R_{pol}$ have been corrected by the Madelung term. The inset shows the decrease with intermolecular distance of the square root of the second moment of the charge density distribution, $\sigma$, indicating a more localized charge distribution around the fullerene.}
\label{fig2}
\end{centering}
\end{figure}

The total energy of the system as a function of the intermolecular distance $R$ and extra charge  
is reported in Figure~\ref{fig2}. 
The total energy for the charged systems has been corrected in order to remove the spurious contribution due to the periodic boundary conditions \cite{MakPay1995} as explained below.

The first  and second minima  of the total energy profile correspond to polymerized (at $R$=$R_{pol}$) and unpolymerized (at larger $R$) molecules, respectively, as shown in the following sections. 
For all the studied anionic states, as well as for the neutral case, the most stable configuration is the polymerized one.
The polymerization  at $R<R_{pol}$  is prevented by an energy barrier which is strongly dependent on the charge of the system.

The sharp energy jump observed in Figure~\ref{fig2} at distance $R \sim R_{pol}$  is a clear indication of the avoided crossing of the two adiabatic surfaces (one for two non bonded C$_{60}$ and the other for the polymerized molecules).
Since the transition state of the polymerization is never observed, the energy barriers here reported are related but do not correspond
 to actual activation energies.

\begin{figure}[ht]
\begin{centering}
\includegraphics[scale=0.65]{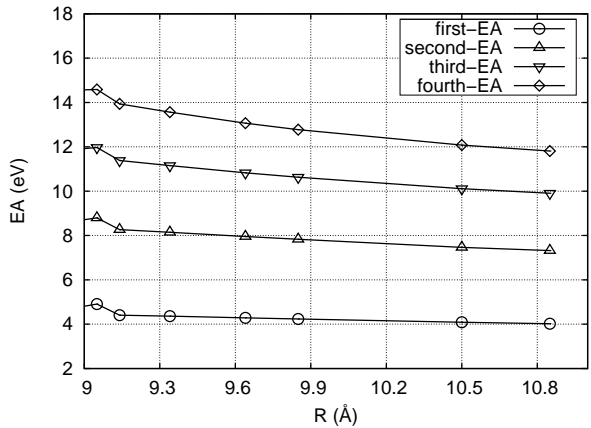}
\caption{Electron affinities for the chain of unpolymerized molecules as a function of the intermolecular distance $R$.}
\label{fig3}
\end{centering}
\end{figure}

\begin{figure}[ht]
\begin{centering}
\includegraphics[scale=0.7]{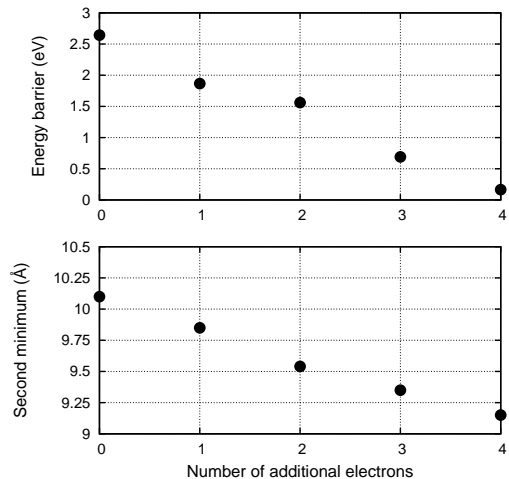}
\vspace*{0.5cm}
\caption{Top panel: decrease of the energy barrier for the [2+2]-cycloaddition reaction as a function of the charge. Bottom panel: decrease of the position of the second minimum of the total energy curve as a function of the charge.}
\label{fig4}
\end{centering}
\end{figure}

In the region of $R>R_{pol}$ we can approximate our system as a purely molecular system and we apply the Madelung energy correction for a lattice of point charges immersed in a neutralizing jellium \cite{MakPay1995}. The energy is then expressed by the following equation:
\begin{equation}
E(L)=E_{LDA}-\frac{q^2\alpha}{2L}+O(L^{-3})
\end{equation}
where $q$ is the extra charge and $\alpha$ is the Madelung constant whose value is known for the simple cubic geometry.
At $R<R_{pol}$, although the ground state of the system is that 
of a polymerized phase, this correction is still a reasonable 
approximation since the localization of the charge between 
the C$_{60}$ molecules is negligible compared to the charge around
 the fullerenes. Also, no discontinuity of the localization of the charge density on the fullerene is observed at $R$=$R_{pol}$ as shown 
in the inset of Figure~\ref{fig2}. The measure of localization 
of the electronic charge on the fullerene is made by computing 
the spherical second moment of the charge density 
distribution, i.e. $\int (r-\langle r\rangle)^2 \rho(r)dr$.
We observe that the total energy of the system drastically decreases  upon addition of extra charge indicating an extremely high positive electron affinity (EA).
 In Figure~\ref{fig4} we report the EAs (from the first-EA to the fourth-EA)  for a linear chain of molecules as a function of the intermolecular distance, for $R>R_{pol}$.
The (first) electron affinity for the two molecule-system is higher than the measured values for the isolated C$_{60}$ (2.67 eV \cite{Yan1987}) and similarly computed values (2.75 eV obtained with GGA by Pederson {\sl et al.} \cite{Ped1992}).
This is consistent with the decrease in quantum confinement of the extra electron in the two molecule-system as compared to the isolated fullerene. 
The monotonic decrease of the electron affinity as a function of the inter-ball distance can be associated as well to an increase in quantum confinement.
 The delocalization of the HOMO wavefunction decreases with increasing intermolecular distance (inset of Figure~\ref{fig2}). 
The second momentum of the charge density distribution at $R$=10.85 \AA\ ($\sim$0.3 \AA) is 30\% smaller than at $R$=9.05 \AA\ ($\sim$0.4 \AA).

The binding energy between neutral C$_{60}$ molecules as obtained by the total energy curve of Figure~\ref{fig2} is 0.15 eV/2 molecules and the equilibrium distance around 10.2 \AA. These values are very similar to those obtained from the Lennard-Jones potential (0.26 eV/2 molecules and 10 \AA, respectively) reported in Ref. \onlinecite{Giri1992} and computed to match experimental values of cohesive energy and lattice constant \cite{Yann1991}. Such agreement can be attributed to the LDA overestimation of binding which implicitly accounts for dispersive forces.

While the equilibrium interaction distance $R_{pol}$  slightly changes as a function of number of electrons added to the system, the 
 energy barrier height is significantly affected by the different charge state.   
The existence of such an energy barrier for the formation of
 covalent bonds between  fullerenes has been already observed 
experimentally \cite{WanHol1994,Dav2001} and  theoretically 
studied \cite{She2007}.
The value of the  energy barrier obtained for the different anionic states is reported in Figure~\ref{fig3}.
 For the neutral system this value (2.64 eV) significantly exceeds the activation energy observed experimentally for photostimulated dimerization (1.25 eV \cite{WanHol1994}) and pressure-induced polymerization  in pristine C$_{60}$ (1.40 eV \cite{Dav2001}) and also the calculated one (1.49 eV in \cite{She2007}). 
This is due to the fact that the energy barriers we show only represent an upper limit to the activation energy.
We observe that upon electron doping the energy 
barrier dramatically decreases from a maximum of 2.64 eV for 
the neutral system down to a minimum of 0.67 eV for a system 
with 4 additional electrons. 
This result proves that the charge actually determines the thermodynamic conditions at which polymerization takes place.
Another important effect of the charge is to decrease the equilibrium intermolecular separation for the unpolymerized phase, as seen from the
charge dependent position of second minimum of the potential energy reported in  Figure~\ref{fig3}. 
We observe a 10\% change in the equilibrium distance between the neutral and $q$=4 case due to the progressive population of bonding states with charge as discussed in the following section.

\subsection{Analysis of the wavefunctions close to the Fermi energy}
\label{wavefunction}

\begin{figure}[ht]
\begin{centering}
\includegraphics[scale=0.4]{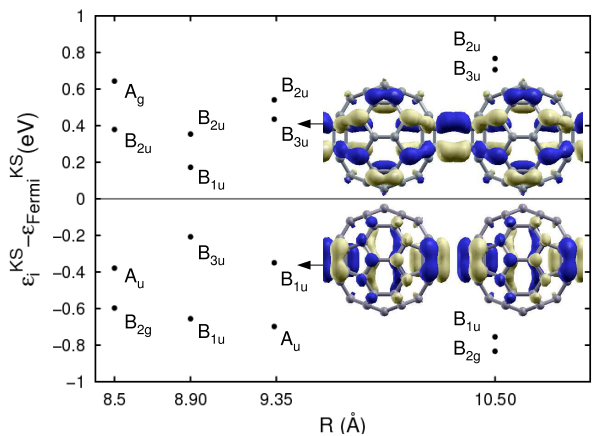}
\caption{Khon-Sham eigenvalues for the neutral system around the Fermi level (indicated by the horizontal line at $\epsilon^{KS}-\epsilon^{KS}_{Fermi}$=0). The states are indicated by
their corresponding symmetry labels (see text) and their energies as shown as a function of the intermolecular distance. Pictures at the inset (color online)
show the density of the bonding (B$_{3u}$) and anti-bonding (B$_{1u}$)  wavefunctions that undergo an occupation inversion at the transition 
distance ($R$=8.90 \AA) }
\label{fig5}
\end{centering}
\end{figure}

The C$_{60}$ molecules in the polymeric chain belong to symmetry point group {\sl D$_{2h}$} and its eigenstates are conveniently classified using the eight irreducible representations of the dihedral point group.  
The bonding and anti-bonding eigenstates relative to the formation of the [2+2] cycloaddition belong to the {\sl B$_{3u}$} and {\sl B$_{1u}$} symmetry, respectively (see Figure~\ref{fig5}).

The eigenfunctions of the C$_{60}$ are strongly modified when the intermolecular distance is changed, preventing us from studying the formation of covalent intermolecular bonds as a function of the distance by a simple analysis of the wavefunction close to the Fermi energy. 
The analysis of the {\sl Mulliken Overlap Population} reported in the following section provides a useful tool for this purpose. 
The analysis of the eigenfunctions shows the existence of four different regions of intermolecular distance where the eigenfunctions of a given symmetry exhibit the same ordering in terms of energy. The result is somewhat different from that previously reported for a C$_{60}$ dimer by Fagerstr\"{o}m and Stafstr\"{o}m \cite{Fage1996}.
Figure~\ref{fig5} shows the  HOMO-1, HOMO, LUMO and LUMO+1 eigenvalues and eigenstate symmetry for $R$=8.5 \AA\/, 8.90 \AA\/, 9.35 \AA\ and 10.85 \AA\ corresponding to $R<R_{pol}$,  $R \sim R_{pol}$, $R>R_{pol}$ and $R \gg R_{pol}$.
 The symmetry of the HOMO and LUMO eigenstates for increasing inter-center separation are  {\sl A$_u$}, {\sl B$_{2u}$}, {\sl B$_{1u}$} and {\sl B$_{2u}$}, {\sl B$_{1u}$}, {\sl B$_{3u}$} respectively. 
The presence of extra electrons results in a little modification of the general picture reported in Figure~\ref{fig5} mostly affecting the eigen energies. 
 This suggests that  for $R<R_{pol}$ the electron doping  would not affect the already formed covalent bonds since the {\sl B$_{2u}$} and {\sl A$_g$} states at higher energy are non-bonding states.
 At $R \sim R_{pol}$ instead, the electron doping would promote the occupation of the {\sl B$_{1u}$} anti-bonding state, making of
this a transition, unstable, state.
At $R>R_{pol}$ and $R \gg R_{pol}$  additional electrons would fill the {\sl B$_{3u}$} bonding state, contributing to favor the covalent bonds between the buckyballs. The photoexcitation of electrons from the anti-bonding {\sl B$_{1u}$} state the bonding {\sl B$_{3u}$} state would also favor the formation of covalent intermolecular bonds.
The filling of this state is the reason for the decrease of optimum intermolecular distance as a function of the extra charge, when the molecules are not yet polymerized.
 It is important to note that at these distances we observe a decrease of the HOMO-LUMO gap  with the increase of $R$ as shown in Figure~\ref{fig5}.

\subsection{Charge Population Analysis}

\begin{figure}[h]
\begin{centering}
\includegraphics[scale=0.75]{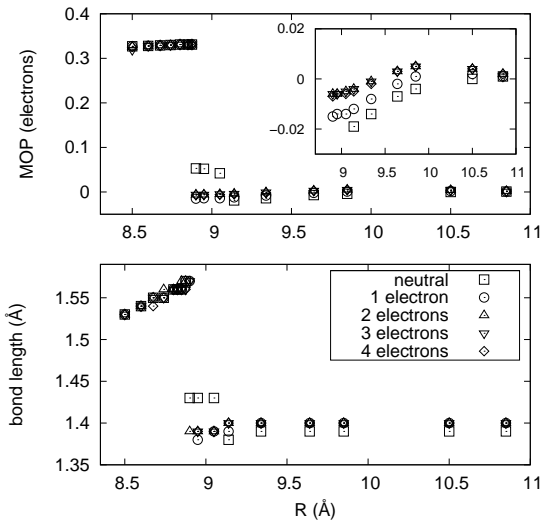}
\vspace*{0.5cm}
\caption{Top panel: evolution as a function of the intermolecular coordinate of the {\sl Mulliken Overlap Population} between atoms 1 and 4 (also 2 and 5, see Figure~\ref{fig1})  participating in the cycloaddition. The inset panel shows the zoom of {\sl MOP} evolution for $R$ $>$ $R_{pol}$. Bottom panel: evolution  of the initial double C-C bond length between atoms 1 and 2 (also 4 and 5).}
\label{fig6}
\end{centering}
\end{figure}

The decomposition of the total charge into atomic contributions in molecules is not a uniquely defined property. 
Nonetheless, bond orders, or non diagonal elements of the density times the overlap matrix product provide useful information about the formation of intermolecular bonds. 
Here we choose to study {\sl Mulliken Overlap Populations} ({\sl MOP}).
 The Mulliken analysis, although strongly dependent on the employed basis set (mostly through the overlap matrix in the product), is a reliable tool in cases where charge considerations are limited to the same atomic species.  
Figure~\ref{fig6} shows the evolution with the intersystem distance and charge of the {\sl MOP}  between atoms 1-4 and 2-5 (see Figure~\ref{fig1}). 
Similarly to the analysis of the wavefunction symmetry, we have identified four regions with a different behavior of the {\sl MOP}.
 At $R<R_{pol}$, the {\sl MOP} is around 80\% of the value observed for the C-C single bonds in the C$_{60}$ molecule, consistent with the formation of covalent intermolecular bonds. 
The addition of electrons to the system does not modify this value because, as discussed in section~\ref{wavefunction}, these occupy  the eigenstates with {\sl B$_{2u}$} and {\sl A$_g$} symmetry corresponding to non-bonding states.
When the center-to-center distance is close to the maximum of the energy barrier (8.88, 8.90 and 8.95 \AA), the {\sl MOP} discontinuously decreases, indicating a  weakening of the intermolecular bonds.
 We observe a value of $\sim$0.05 electrons for the neutral system while smaller values are found for the charged system. 
This is due to the occupation of the anti-bonding {\sl B$_{1u}$} eigenstate.
At bigger intermolecular distances, at  $R>R_{pol}$, the polymerization does not occur since the  {\sl MOP} value drops to negative values indicating a repulsive overlap. 
 The progressive occupation of the {\sl B$_{3u}$} bonding state by 1 and 2 electrons progressively increases the bond order of the cycloadduct and a positive overlap is observed (inset of Figure~\ref{fig6}). 
The {\sl MOP} value remains constant upon occupation of the {\sl B$_{2u}$} state  which is a non-bonding state whithin the four membered ring but it has a bonding character between atoms 3 and 6 and the other four equivalent pairs of atoms (see Figure~\ref{fig1}). 
This is the reason for the almost linear decrease of the second equilibrium intermolecular minimum for the unpolymerized fullerenes shown in Figure~\ref{fig2}. A progressive occupation of bonding states between the molecules with increasing number of extra electrons will favor shorter intermolecular 
distances and promote the [2+2]-cycloaddition reaction at lower pressures.

\subsection{Structural Analysis}
 Both buckyballs contained in the supercell are subject to an identical
 structural change with distance and electron doping compatible with the {\sl D$_{2h}$ }symmetry.
Such a change  mostly affects the ten atoms 
belonging to the 2 adjacent hexagons 
facing the neighboring molecule as shown in Figure~\ref{fig6} and previously reported \cite{She2007}.
 The bond length of the single bonds involving the atoms  
participating in the cycloaddition (between atoms 1-4 and 2-5  in Figure~\ref{fig1}) slightly changes as
 function of distance and charge (up to a maximum of 0.08 \AA). 
On the other side, the initial double C-C bonds (between atoms 1-2 and 4-5) drastically
 change with distance (up to a maximum of 0.20 \AA) and slightly with charge as reported in Figure~\ref{fig6}.
The result of the structural analysis is in agreement with the wavefunction and {\sl MOP} analysis.
 At $R<R_{pol}$, we observe a lengthening of the double C-C bond  compared to the isolated molecule, up to a 10\% increase,
consistent with the formation of a [2+2]-cycloadduct.
 At these short intermolecular distances, the C-C bond between atoms 1-2 is not affected by the addition of electrons. 
  At $R>R_{pol}$ instead, this C-C bond length decreases abruptly as observed for the {\sl MOP}. In particular, it recovers the same value as in the isolated molecule showing that at these distances no polymerization of the molecules takes place. 
The addition of 1-2 electrons has a minor effect on the C-C bond length consistent with the {\sl MOP} and wavefunction analysis while the presence of 3-4 electrons does not further change this value. Instead, the C-C bond length between atoms 3-6 is reduced in agreement with the wavefunction analysis.

\section{Conclusions}

We have performed {\em ab initio} DFT calculations
 of C$_{60}$ molecular chains in order to
 understand interplay between charge doping and intermolecular separation on the polymerization process of the chains.

We have studied the potential energy surface, the symmetry of the wavefunctions around the Fermi level,  the {\sl MOP} between atoms of the cycloadduct  and the evolution of initial C-C double bonding length.  

The analysis of the potential energy surface shows that the total energy profile as a function of the intermolecular distance has a double-well shape.
 The {\sl MOP} and structural analysis show that the first and second minimum of the total energy profile correspond to polymerized and unpolymerized molecules, respectively. 
 The equilibrium intermolecular distance of the polymerized C$_{60}$ is not significantly shifted with charge doping.
 The [2+2]-cycloaddition reaction is prevented by an energy barrier  which is related to the activation energy. 
The addition of electrons to the system results in an almost linear decrease of the energy barrier as a function of the charge consistent with the study reported in Ref.~\onlinecite{Fage1996}.
 This is a very important result because it proves that in charged fullerenes the polymerization can occur at lower temperature.
 This is also consistent with the computational study performed by Sheka \cite{She2007} and the 
experimental evidence of a spontaneous one-dimensional polymerization for 
the alkali-intercalated fullerenes MC$_{60}$ (M=K, Rb and Cs) under reversible
 solid-state transformation from the high-temperature phase \cite{Ste1994,LauMor1998,Ver2002}.

 The equilibrium intermolecular distance of the unpolymerized molecules is appreciably affected by the electron doping.
 In particular, the unpolymerized equilibrium distance  linearly decreases due to the progressive  occupation of bonding wavefunctions.

We conclude that polymerization depends on both the center-to-center distance of fullerenes and the negative doping of the system.
 Experimentally, pressure and temperature are required to approach the molecules and overcome the energy barrier, respectively.
 A different electron doping would affect the height of the energy barrier but not the distance at which polymerization occurs. 
In particular, we observed that up to a doping of 4 electrons per 2 molecules, the energy barrier progressively decreases. 
%
%
The size of the alkali cation clearly affects the intermolecular distance but also the charge transfer to the C$_{60}$ due to the decrease of the ionization potential with increasing ionic size. The potential energy curve suggests that if the intermolecular distance is $R>R_{pol}$, for a given number of metal atoms, bigger size cations should favor the polymerization due to the larger decrease of the energy barrier observed with higher electron doping.

\section{Acknowledgments}
We acknowledge discussions with Xavier Blase, Alberto Garc\'{i}a, Enric Canadell and Berend Smit. MVFS was funded by DOE award number DE-SC0003871. The work at the University of California were supported as part of the Center for Gas Separations Relevant to Clean Energy Technologies, an Energy Frontier Research Center funded by the U.S. Department of Energy, Office of Science,
Office of Basic Energy Sciences under Award Number DE-SC0001015.

\onecolumngrid

\begin{thebibliography}{36}
\expandafter\ifx\csname natexlab\endcsname\relax\def\natexlab#1{#1}\fi
\expandafter\ifx\csname bibnamefont\endcsname\relax
  \def\bibnamefont#1{#1}\fi
\expandafter\ifx\csname bibfnamefont\endcsname\relax
  \def\bibfnamefont#1{#1}\fi
\expandafter\ifx\csname citenamefont\endcsname\relax
  \def\citenamefont#1{#1}\fi
\expandafter\ifx\csname url\endcsname\relax
  \def\url#1{\texttt{#1}}\fi
\expandafter\ifx\csname urlprefix\endcsname\relax\def\urlprefix{URL }\fi
\providecommand{\bibinfo}[2]{#2}
\providecommand{\eprint}[2][]{\url{#2}}

\bibitem[{\citenamefont{Rao et~al.}(1993)\citenamefont{Rao, Zhou, Wang, Hager,
  Holden, Wang, Lee, Bi, Eklund, Cornett et~al.}}]{Rao1993}
\bibinfo{author}{\bibfnamefont{A.~M.} \bibnamefont{Rao}},
  \bibinfo{author}{\bibfnamefont{P.}~\bibnamefont{Zhou}},
  \bibinfo{author}{\bibfnamefont{K.~A.} \bibnamefont{Wang}},
  \bibinfo{author}{\bibfnamefont{G.~T.} \bibnamefont{Hager}},
  \bibinfo{author}{\bibfnamefont{J.~M.} \bibnamefont{Holden}},
  \bibinfo{author}{\bibfnamefont{Y.}~\bibnamefont{Wang}},
  \bibinfo{author}{\bibfnamefont{W.~T.} \bibnamefont{Lee}},
  \bibinfo{author}{\bibfnamefont{X.~X.} \bibnamefont{Bi}},
  \bibinfo{author}{\bibfnamefont{P.~C.} \bibnamefont{Eklund}},
  \bibinfo{author}{\bibfnamefont{D.~S.} \bibnamefont{Cornett}},
  \bibnamefont{et~al.}, \bibinfo{journal}{Science}
  \textbf{\bibinfo{volume}{259}}, \bibinfo{pages}{955} (\bibinfo{year}{1993}).

\bibitem[{\citenamefont{Iwasa et~al.}(1994)\citenamefont{Iwasa, Arima, Fleming,
  Siegrist, Zhou, Haddon, Rothberg, Lyons, Jr., Hebard et~al.}}]{Iwa1994}
\bibinfo{author}{\bibfnamefont{Y.}~\bibnamefont{Iwasa}},
  \bibinfo{author}{\bibfnamefont{T.}~\bibnamefont{Arima}},
  \bibinfo{author}{\bibfnamefont{R.~M.} \bibnamefont{Fleming}},
  \bibinfo{author}{\bibfnamefont{T.}~\bibnamefont{Siegrist}},
  \bibinfo{author}{\bibfnamefont{O.}~\bibnamefont{Zhou}},
  \bibinfo{author}{\bibfnamefont{R.~C.} \bibnamefont{Haddon}},
  \bibinfo{author}{\bibfnamefont{L.~J.} \bibnamefont{Rothberg}},
  \bibinfo{author}{\bibfnamefont{K.~B.} \bibnamefont{Lyons}},
  \bibinfo{author}{\bibfnamefont{H.~L.~C.} \bibnamefont{Jr.}},
  \bibinfo{author}{\bibfnamefont{A.~F.} \bibnamefont{Hebard}},
  \bibnamefont{et~al.}, \bibinfo{journal}{Science}
  \textbf{\bibinfo{volume}{264}}, \bibinfo{pages}{1570} (\bibinfo{year}{1994}).

\bibitem[{\citenamefont{{N\'u\~nez-Regueiro}
  et~al.}(1995)\citenamefont{{N\'u\~nez-Regueiro}, Marques, Hodeau, B\`ethoux,
  and Perroux}}]{Nun1995}
\bibinfo{author}{\bibfnamefont{M.}~\bibnamefont{{N\'u\~nez-Regueiro}}},
  \bibinfo{author}{\bibfnamefont{L.}~\bibnamefont{Marques}},
  \bibinfo{author}{\bibfnamefont{J.-L.} \bibnamefont{Hodeau}},
  \bibinfo{author}{\bibfnamefont{O.}~\bibnamefont{B\`ethoux}},
  \bibnamefont{and} \bibinfo{author}{\bibfnamefont{M.}~\bibnamefont{Perroux}},
  \bibinfo{journal}{Phys. Rev. Lett.} \textbf{\bibinfo{volume}{74}},
  \bibinfo{pages}{278} (\bibinfo{year}{1995}).

\bibitem[{\citenamefont{Sundqvist}(1999)}]{Sun1999}
\bibinfo{author}{\bibfnamefont{B.}~\bibnamefont{Sundqvist}},
  \bibinfo{journal}{Adv. in Phys.} \textbf{\bibinfo{volume}{48}},
  \bibinfo{pages}{1} (\bibinfo{year}{1999}).

\bibitem[{\citenamefont{Zhao et~al.}(1994)\citenamefont{Zhao, Poirier, Pechman,
  and Weaver}}]{Zhao1994}
\bibinfo{author}{\bibfnamefont{I.~B.} \bibnamefont{Zhao}},
  \bibinfo{author}{\bibfnamefont{D.~N.} \bibnamefont{Poirier}},
  \bibinfo{author}{\bibfnamefont{R.~J.} \bibnamefont{Pechman}},
  \bibnamefont{and} \bibinfo{author}{\bibfnamefont{J.~H.}
  \bibnamefont{Weaver}}, \bibinfo{journal}{Appl. Phys. Lett.}
  \textbf{\bibinfo{volume}{64}}, \bibinfo{pages}{577} (\bibinfo{year}{1994}).

\bibitem[{\citenamefont{Pekker et~al.}(1994)\citenamefont{Pekker, Janossy,
  Mihali, Chauvet, and Forro}}]{Pek1994}
\bibinfo{author}{\bibfnamefont{S.}~\bibnamefont{Pekker}},
  \bibinfo{author}{\bibfnamefont{A.}~\bibnamefont{Janossy}},
  \bibinfo{author}{\bibfnamefont{L.}~\bibnamefont{Mihali}},
  \bibinfo{author}{\bibfnamefont{O.}~\bibnamefont{Chauvet}}, \bibnamefont{and}
  \bibinfo{author}{\bibfnamefont{L.}~\bibnamefont{Forro}},
  \bibinfo{journal}{Science} \textbf{\bibinfo{volume}{265}},
  \bibinfo{pages}{1077} (\bibinfo{year}{1994}).

\bibitem[{\citenamefont{Stephens et~al.}(1994)\citenamefont{Stephens, Faigel,
  Tegze, Janossy, Pekker, Oszlanyi, and Forro}}]{Ste1994}
\bibinfo{author}{\bibfnamefont{P.~W.} \bibnamefont{Stephens}},
  \bibinfo{author}{\bibfnamefont{G.~B.~G.} \bibnamefont{Faigel}},
  \bibinfo{author}{\bibfnamefont{M.}~\bibnamefont{Tegze}},
  \bibinfo{author}{\bibfnamefont{A.}~\bibnamefont{Janossy}},
  \bibinfo{author}{\bibfnamefont{S.}~\bibnamefont{Pekker}},
  \bibinfo{author}{\bibfnamefont{G.}~\bibnamefont{Oszlanyi}}, \bibnamefont{and}
  \bibinfo{author}{\bibfnamefont{L.}~\bibnamefont{Forro}},
  \bibinfo{journal}{Nature} \textbf{\bibinfo{volume}{370}},
  \bibinfo{pages}{636} (\bibinfo{year}{1994}).

\bibitem[{\citenamefont{Launois et~al.}(1998)\citenamefont{Launois, Moret,
  Hone, and Zetti}}]{LauMor1998}
\bibinfo{author}{\bibfnamefont{P.}~\bibnamefont{Launois}},
  \bibinfo{author}{\bibfnamefont{R.}~\bibnamefont{Moret}},
  \bibinfo{author}{\bibfnamefont{J.}~\bibnamefont{Hone}}, \bibnamefont{and}
  \bibinfo{author}{\bibfnamefont{A.}~\bibnamefont{Zettl}},
  \bibinfo{journal}{Physical Review Letters} \textbf{\bibinfo{volume}{81}},
  \bibinfo{pages}{4420} (\bibinfo{year}{1998}).

\bibitem[{\citenamefont{Verberck et~al.}(2002)\citenamefont{Verberck, Michel,
  and Nikolaev}}]{Ver2002}
\bibinfo{author}{\bibfnamefont{B.}~\bibnamefont{Verberck}},
  \bibinfo{author}{\bibfnamefont{H.}~\bibnamefont{Michel}}, \bibnamefont{and}
  \bibinfo{author}{\bibfnamefont{A.~V.} \bibnamefont{Nikolaev}},
  \bibinfo{journal}{J. Chem Phys.} \textbf{\bibinfo{volume}{116}},
  \bibinfo{pages}{10462} (\bibinfo{year}{2002}).

\bibitem[{\citenamefont{Dresselhaus et~al.}(1996)\citenamefont{Dresselhaus,
  Dresselhaus, and Eklund}}]{Dre1996}
\bibinfo{author}{\bibfnamefont{M.~S.} \bibnamefont{Dresselhaus}},
  \bibinfo{author}{\bibfnamefont{G.}~\bibnamefont{Dresselhaus}},
  \bibnamefont{and} \bibinfo{author}{\bibfnamefont{P.~C.}
  \bibnamefont{Eklund}}, \bibinfo{journal}{Science of Fullerenes and Carbon
  Nanotubes (Academic Press Inc.)}  (\bibinfo{year}{1996}).

\bibitem[{\citenamefont{Margadonna et~al.}(2004)\citenamefont{Margadonna,
  Pontiroli, Belli, T.~Shikora, and Brunelli}}]{Mar2004}
\bibinfo{author}{\bibfnamefont{S.}~\bibnamefont{Margadonna}},
  \bibinfo{author}{\bibfnamefont{D.}~\bibnamefont{Pontiroli}},
  \bibinfo{author}{\bibfnamefont{M.}~\bibnamefont{Belli}},
  \bibinfo{author}{\bibfnamefont{M.~R.} \bibnamefont{T.~Shikora}},
  \bibnamefont{and} \bibinfo{author}{\bibfnamefont{M.}~\bibnamefont{Brunelli}},
  \bibinfo{journal}{J. A. Chem. Soc.} \textbf{\bibinfo{volume}{126}},
  \bibinfo{pages}{15032} (\bibinfo{year}{2004}).

\bibitem[{\citenamefont{Halevi}(2002)}]{Emi2002}
\bibinfo{author}{\bibfnamefont{E.~A.} \bibnamefont{Halevi}},
  \bibinfo{journal}{J. Phys. Org. Chem.} \textbf{\bibinfo{volume}{15}},
  \bibinfo{pages}{519} (\bibinfo{year}{2002}).

\bibitem[{\citenamefont{Poloni et~al.}(2008{\natexlab{a}})\citenamefont{Poloni,
  Machon, Fernandez-Serra, {Le Floch}, Pascarelli, Montagnac, Cardon, and
  {San-Miguel}}}]{PolB2008}
\bibinfo{author}{\bibfnamefont{R.}~\bibnamefont{Poloni}},
  \bibinfo{author}{\bibfnamefont{D.}~\bibnamefont{Machon}},
  \bibinfo{author}{\bibfnamefont{M.~V.} \bibnamefont{Fernandez-Serra}},
  \bibinfo{author}{\bibfnamefont{S.}~\bibnamefont{{Le Floch}}},
  \bibinfo{author}{\bibfnamefont{S.}~\bibnamefont{Pascarelli}},
  \bibinfo{author}{\bibfnamefont{G.}~\bibnamefont{Montagnac}},
  \bibinfo{author}{\bibfnamefont{H.}~\bibnamefont{Cardon}}, \bibnamefont{and}
  \bibinfo{author}{\bibfnamefont{A.}~\bibnamefont{{San-Miguel}}},
  \bibinfo{journal}{Phys. Rev. B} \textbf{\bibinfo{volume}{77}},
  \bibinfo{pages}{125413} (\bibinfo{year}{2008}{\natexlab{a}}).

\bibitem[{\citenamefont{Poloni et~al.}(2008{\natexlab{b}})\citenamefont{Poloni,
  Aquilanti, Toulemonde, Pascarelli, {Le Floch}, Machon, {Martinez-Blanco},
  Morard, and {San-Miguel}}}]{PolC2008}
\bibinfo{author}{\bibfnamefont{R.}~\bibnamefont{Poloni}},
  \bibinfo{author}{\bibfnamefont{G.}~\bibnamefont{Aquilanti}},
  \bibinfo{author}{\bibfnamefont{P.}~\bibnamefont{Toulemonde}},
  \bibinfo{author}{\bibfnamefont{S.}~\bibnamefont{Pascarelli}},
  \bibinfo{author}{\bibfnamefont{S.}~\bibnamefont{{Le Floch}}},
  \bibinfo{author}{\bibfnamefont{D.}~\bibnamefont{Machon}},
  \bibinfo{author}{\bibfnamefont{D.}~\bibnamefont{{Martinez-Blanco}}},
  \bibinfo{author}{\bibfnamefont{G.}~\bibnamefont{Morard}}, \bibnamefont{and}
  \bibinfo{author}{\bibfnamefont{A.}~\bibnamefont{{San-Miguel}}},
  \bibinfo{journal}{Phys. Rev. B} \textbf{\bibinfo{volume}{77}},
  \bibinfo{pages}{205433} (\bibinfo{year}{2008}{\natexlab{b}}).

\bibitem[{\citenamefont{Poloni et~al.}(2009)\citenamefont{Poloni, Toulemonde,
  Machon, {Le Floch}, Pascarelli, and {San-Miguel}}}]{PolTou2010}
\bibinfo{author}{\bibfnamefont{R.}~\bibnamefont{Poloni}},
  \bibinfo{author}{\bibfnamefont{P.}~\bibnamefont{Toulemonde}},
  \bibinfo{author}{\bibfnamefont{D.}~\bibnamefont{Machon}},
  \bibinfo{author}{\bibfnamefont{S.}~\bibnamefont{{Le Floch}}},
  \bibinfo{author}{\bibfnamefont{S.}~\bibnamefont{Pascarelli}},
  \bibnamefont{and}
  \bibinfo{author}{\bibfnamefont{A.}~\bibnamefont{{San-Miguel}}},
  \bibinfo{journal}{High Press. Res.} \textbf{\bibinfo{volume}{29}},
  \bibinfo{pages}{108} (\bibinfo{year}{2009}).

\bibitem[{\citenamefont{Pederson and Quong}(1992)}]{Ped1992}
\bibinfo{author}{\bibfnamefont{M.~R.} \bibnamefont{Pederson}} \bibnamefont{and}
  \bibinfo{author}{\bibfnamefont{A.~A.} \bibnamefont{Quong}},
  \bibinfo{journal}{Phys. Rev. B} \textbf{\bibinfo{volume}{46}},
  \bibinfo{pages}{13584} (\bibinfo{year}{1992}).

\bibitem[{\citenamefont{Santos et~al.}(2003)\citenamefont{Santos, Longo, and
  Taft}}]{SanLon2003}
\bibinfo{author}{\bibfnamefont{J.~D.} \bibnamefont{Santos}},
  \bibinfo{author}{\bibfnamefont{E.}~\bibnamefont{Longo}}, \bibnamefont{and}
  \bibinfo{author}{\bibfnamefont{C.~A.} \bibnamefont{Taft}},
  \bibinfo{journal}{Theo. Chem.} \textbf{\bibinfo{volume}{625}},
  \bibinfo{pages}{189} (\bibinfo{year}{2003}).

\bibitem[{\citenamefont{K\"{u}rti and N\'{e}meth}(1996)}]{Kurti1996}
\bibinfo{author}{\bibfnamefont{J.}~\bibnamefont{K\"{u}rti}} \bibnamefont{and}
  \bibinfo{author}{\bibfnamefont{K.}~\bibnamefont{N\'{e}meth}},
  \bibinfo{journal}{Chem Phys. Lett.} \textbf{\bibinfo{volume}{256}},
  \bibinfo{pages}{119} (\bibinfo{year}{1996}).

\bibitem[{\citenamefont{Fagerstr\"{o}m and Stafstr\"{o}m}(1996)}]{Fage1996}
\bibinfo{author}{\bibfnamefont{J.}~\bibnamefont{Fagerstr\"{o}m}}
  \bibnamefont{and}
  \bibinfo{author}{\bibfnamefont{S.}~\bibnamefont{Stafstr\"{o}m}},
  \bibinfo{journal}{Phys. Rev. B} \textbf{\bibinfo{volume}{53}},
  \bibinfo{pages}{13150} (\bibinfo{year}{1996}).

\bibitem[{\citenamefont{Sheka}(2007)}]{She2007}
\bibinfo{author}{\bibfnamefont{E.~F.} \bibnamefont{Sheka}},
  \bibinfo{journal}{Chem. Phys. Lett.} \textbf{\bibinfo{volume}{438}},
  \bibinfo{pages}{119} (\bibinfo{year}{2007}).

\bibitem[{\citenamefont{Miguel}(2006)}]{SanMiguel2006}
\bibinfo{author}{\bibfnamefont{A.~S.} \bibnamefont{Miguel}},
  \bibinfo{journal}{Chem. Soc. Rev.} \textbf{\bibinfo{volume}{35}},
  \bibinfo{pages}{876} (\bibinfo{year}{2006}).

\bibitem[{\citenamefont{Soler et~al.}(2002)\citenamefont{Soler, Artacho, Gale,
  Garc\'ia, Junquera, Ordej\'on, and Sanchez-Portal}}]{SolArt2002}
\bibinfo{author}{\bibfnamefont{J.~M.} \bibnamefont{Soler}},
  \bibinfo{author}{\bibfnamefont{E.}~\bibnamefont{Artacho}},
  \bibinfo{author}{\bibfnamefont{J.~D.} \bibnamefont{Gale}},
  \bibinfo{author}{\bibfnamefont{A.}~\bibnamefont{Garc\'ia}},
  \bibinfo{author}{\bibfnamefont{J.}~\bibnamefont{Junquera}},
  \bibinfo{author}{\bibfnamefont{P.}~\bibnamefont{Ordej\'on}},
  \bibnamefont{and} \bibinfo{author}{\bibfnamefont{D.~S.}
  \bibnamefont{Sanchez-Portal}}, \bibinfo{journal}{J. Phys: Condens. Matter}
  \textbf{\bibinfo{volume}{14}}, \bibinfo{pages}{2745} (\bibinfo{year}{2002}).

\bibitem[{\citenamefont{Perdew and Zunger}(1981)}]{PerZun1981}
\bibinfo{author}{\bibfnamefont{J.~P.} \bibnamefont{Perdew}} \bibnamefont{and}
  \bibinfo{author}{\bibfnamefont{A.}~\bibnamefont{Zunger}},
  \bibinfo{journal}{Phys. Rev. B} \textbf{\bibinfo{volume}{23}},
  \bibinfo{pages}{5048} (\bibinfo{year}{1981}).

\bibitem[{\citenamefont{Girard et~al.}(1994)\citenamefont{Girard, Lambin,
  Dereux, and Lucas}}]{GirLam1994}
\bibinfo{author}{\bibfnamefont{C.}~\bibnamefont{Girard}},
  \bibinfo{author}{\bibfnamefont{P.}~\bibnamefont{Lambin}},
  \bibinfo{author}{\bibfnamefont{A.}~\bibnamefont{Dereux}}, \bibnamefont{and}
  \bibinfo{author}{\bibfnamefont{A.~A.} \bibnamefont{Lucas}},
  \bibinfo{journal}{Phys. Rev. B} \textbf{\bibinfo{volume}{49}},
  \bibinfo{pages}{11425} (\bibinfo{year}{1994}).

\bibitem[{\citenamefont{Andersson and Rydberg}(1999)}]{AndRyd1999}
\bibinfo{author}{\bibfnamefont{Y.}~\bibnamefont{Andersson}} \bibnamefont{and}
  \bibinfo{author}{\bibfnamefont{H.}~\bibnamefont{Rydberg}},
  \bibinfo{journal}{Physica Scripta} \textbf{\bibinfo{volume}{60}},
  \bibinfo{pages}{211} (\bibinfo{year}{1999}).

\bibitem[{\citenamefont{Junquera et~al.}(2001)\citenamefont{Junquera, Paz,
  S\'anchez-Portal, and Artacho}}]{Jun2001}
\bibinfo{author}{\bibfnamefont{J.}~\bibnamefont{Junquera}},
  \bibinfo{author}{\bibfnamefont{O.}~\bibnamefont{Paz}},
  \bibinfo{author}{\bibfnamefont{D.}~\bibnamefont{S\'anchez-Portal}},
  \bibnamefont{and} \bibinfo{author}{\bibfnamefont{E.}~\bibnamefont{Artacho}},
  \bibinfo{journal}{Phys. Rev. B} \textbf{\bibinfo{volume}{64}},
  \bibinfo{pages}{235111} (\bibinfo{year}{2001}).

\bibitem[{\citenamefont{Anglada et~al.}(2002)\citenamefont{Anglada, Soler,
  Junquera, and Artacho}}]{Ang2002}
\bibinfo{author}{\bibfnamefont{E.}~\bibnamefont{Anglada}},
  \bibinfo{author}{\bibfnamefont{J.~M.} \bibnamefont{Soler}},
  \bibinfo{author}{\bibfnamefont{J.}~\bibnamefont{Junquera}}, \bibnamefont{and}
  \bibinfo{author}{\bibfnamefont{E.}~\bibnamefont{Artacho}},
  \bibinfo{journal}{Phys. Rev. B} \textbf{\bibinfo{volume}{66}},
  \bibinfo{pages}{205101} (\bibinfo{year}{2002}).

\bibitem[{\citenamefont{Leslie and Gillan}(1985)}]{LesGil1985}
\bibinfo{author}{\bibfnamefont{M.}~\bibnamefont{Leslie}} \bibnamefont{and}
  \bibinfo{author}{\bibfnamefont{M.~J.} \bibnamefont{Gillan}},
  \bibinfo{journal}{J. Phys. C} \textbf{\bibinfo{volume}{18}},
  \bibinfo{pages}{973} (\bibinfo{year}{1985}).

\bibitem[{\citenamefont{{De Vita} et~al.}(1992)\citenamefont{{De Vita}, Gillan,
  Lin, Payne, Stich, and Clarke}}]{DeVGil1992}
\bibinfo{author}{\bibfnamefont{A.}~\bibnamefont{{De Vita}}},
  \bibinfo{author}{\bibfnamefont{M.~J.} \bibnamefont{Gillan}},
  \bibinfo{author}{\bibfnamefont{J.~S.} \bibnamefont{Lin}},
  \bibinfo{author}{\bibfnamefont{M.~C.} \bibnamefont{Payne}},
  \bibinfo{author}{\bibfnamefont{I.}~\bibnamefont{Stich}}, \bibnamefont{and}
  \bibinfo{author}{\bibfnamefont{L.~J.} \bibnamefont{Clarke}},
  \bibinfo{journal}{Phys. Rev. B} \textbf{\bibinfo{volume}{46}},
  \bibinfo{pages}{12964} (\bibinfo{year}{1992}).

\bibitem[{\citenamefont{Moret et~al.}(1997)\citenamefont{Moret, Launois,
  Persson, and Sundqvist}}]{MorLau1997}
\bibinfo{author}{\bibfnamefont{R.}~\bibnamefont{Moret}},
  \bibinfo{author}{\bibfnamefont{P.}~\bibnamefont{Launois}},
  \bibinfo{author}{\bibfnamefont{P.~A.} \bibnamefont{Persson}},
  \bibnamefont{and}
  \bibinfo{author}{\bibfnamefont{B.}~\bibnamefont{Sundqvist}},
  \bibinfo{journal}{Europhys. Lett.} \textbf{\bibinfo{volume}{40}},
  \bibinfo{pages}{55} (\bibinfo{year}{1997}).

\bibitem[{\citenamefont{Makov and Payne}(1995)}]{MakPay1995}
\bibinfo{author}{\bibfnamefont{G.}~\bibnamefont{Makov}} \bibnamefont{and}
  \bibinfo{author}{\bibfnamefont{M.~C.} \bibnamefont{Payne}},
  \bibinfo{journal}{Phys. Rev. B} \textbf{\bibinfo{volume}{51}},
  \bibinfo{pages}{4014} (\bibinfo{year}{1995}).

\bibitem[{\citenamefont{Yang et~al.}(1987)\citenamefont{Yang, Pettiette,
  Conceicao, Cheshnovsky, and Smalley}}]{Yan1987}
\bibinfo{author}{\bibfnamefont{S.~H.} \bibnamefont{Yang}},
  \bibinfo{author}{\bibfnamefont{C.~L.} \bibnamefont{Pettiette}},
  \bibinfo{author}{\bibfnamefont{J.}~\bibnamefont{Conceicao}},
  \bibinfo{author}{\bibfnamefont{O.}~\bibnamefont{Cheshnovsky}},
  \bibnamefont{and} \bibinfo{author}{\bibfnamefont{R.~E.}
  \bibnamefont{Smalley}}, \bibinfo{journal}{Chem. Phys. Lett.}
  \textbf{\bibinfo{volume}{139}}, \bibinfo{pages}{223} (\bibinfo{year}{1987}).

\bibitem[{\citenamefont{Girifalco}(1992)}]{Giri1992}
\bibinfo{author}{\bibfnamefont{L.~A.} \bibnamefont{Girifalco}},
  \bibinfo{journal}{J. Phys. Chem.} \textbf{\bibinfo{volume}{96}},
  \bibinfo{pages}{858} (\bibinfo{year}{1992}).

\bibitem[{\citenamefont{Yannoni et~al.}(1991)\citenamefont{Yannoni, Johnson,
  Meijer, Bethune, and Salem}}]{Yann1991}
\bibinfo{author}{\bibfnamefont{C.~S.} \bibnamefont{Yannoni}},
  \bibinfo{author}{\bibfnamefont{R.~D.} \bibnamefont{Johnson}},
  \bibinfo{author}{\bibfnamefont{G.}~\bibnamefont{Meijer}},
  \bibinfo{author}{\bibfnamefont{D.~S.} \bibnamefont{Bethune}},
  \bibnamefont{and} \bibinfo{author}{\bibfnamefont{J.~R.} \bibnamefont{Salem}},
  \bibinfo{journal}{J. Phys. Chem.} \textbf{\bibinfo{volume}{95}},
  \bibinfo{pages}{9} (\bibinfo{year}{1991}).

\bibitem[{\citenamefont{Wang et~al.}(1994)\citenamefont{Wang, Holden, Bi, and
  Eklund}}]{WanHol1994}
\bibinfo{author}{\bibfnamefont{Y.}~\bibnamefont{Wang}},
  \bibinfo{author}{\bibfnamefont{J.~M.} \bibnamefont{Holden}},
  \bibinfo{author}{\bibfnamefont{X.~X.} \bibnamefont{Bi}}, \bibnamefont{and}
  \bibinfo{author}{\bibfnamefont{P.~C.} \bibnamefont{Eklund}},
  \bibinfo{journal}{Chem. Phys. Lett.} \textbf{\bibinfo{volume}{217}},
  \bibinfo{pages}{413} (\bibinfo{year}{1994}).

\bibitem[{\citenamefont{Davydov et~al.}(2001)\citenamefont{Davydov,
  Kashevarova, Rakhmanina, Senyavin, Pronina, Oleynikoy, Agafonov, Ceolin,
  Allouchi, and Szwarc}}]{Dav2001}
\bibinfo{author}{\bibfnamefont{V.~A.} \bibnamefont{Davydov}},
  \bibinfo{author}{\bibfnamefont{L.~S.} \bibnamefont{Kashevarova}},
  \bibinfo{author}{\bibfnamefont{A.~V.} \bibnamefont{Rakhmanina}},
  \bibinfo{author}{\bibfnamefont{V.~M.} \bibnamefont{Senyavin}},
  \bibinfo{author}{\bibfnamefont{O.~P.} \bibnamefont{Pronina}},
  \bibinfo{author}{\bibfnamefont{N.~N.} \bibnamefont{Oleynikoy}},
  \bibinfo{author}{\bibfnamefont{V.}~\bibnamefont{Agafonov}},
  \bibinfo{author}{\bibfnamefont{R.}~\bibnamefont{Ceolin}},
  \bibinfo{author}{\bibfnamefont{H.}~\bibnamefont{Allouchi}}, \bibnamefont{and}
  \bibinfo{author}{\bibfnamefont{H.}~\bibnamefont{Szwarc}},
  \bibinfo{journal}{Chem. Phys. Lett.} \textbf{\bibinfo{volume}{333}},
  \bibinfo{pages}{224} (\bibinfo{year}{2001}).



\end{thebibliography}
\end{document}